\documentclass[12pt,preprint]{aastex}

\def\msun{{\rm M$_{\odot}$}}
\def\rsun{{\rm R$_{\odot}$}}

\begin{document}

\title{Simulations of Stellar Collisions Involving Pre-Main Sequence Stars}

\author{Daniel Laycock \& Alison Sills}
\affil{Department of Physics and Astronomy, McMaster University, 1280 Main Street West, Hamilton, Ontario, L8S 4M1, Canada}
\email{laycocdt@mcmaster.ca, asills@mcmaster.ca}

\begin{abstract}
In this paper, we present the results of smoothed particle
hydrodynamic (SPH) simulations of collisions between pre-main sequence
stars and a variety of other kinds of stars.  Simulations over a range
of impact parameters and velocities were performed.  We find that
pre-main sequence stars tend to ``wrap themselves'' around their
impactor.  We discuss the probable evolutionary state of products of
collisions between pre-main sequence stars and pre-main sequence, main
sequence, giant branch, and compact stars. The nature of the collision
product does not depend strongly on the impact parameter or the
velocity of the collision.
\end{abstract}


\section{Introduction}
Over the past decade, dynamical models of stellar clusters have become
much more realistic. This realism takes the form of an increasingly
complicated treatment of the individual stars in the cluster. For
years, simple dynamical models only considered stars as single
equal-mass, non-evolving points \citep[e.g.][]{C80}. The introduction
of a mass function into dynamical models quickly necessitated some
treatment of stellar evolution, since high mass stars have much
shorter lifetimes than low mass stars. Mass loss from high mass stars
can also have a substantial impact on the dynamical evolution of the
cluster \citep{EHI87}. Binary stars also have a substantial impact on
the cluster, by acting as energy sources or sinks. Before globular
clusters were thought to have primordial binaries, dynamically
produced binaries were recognized as a key population for halting core
collapse \citep{Hetal92}. Clusters also have primordial binaries, and
those binary systems can affect the cluster evolution from its birth.

Stellar dynamicists also realized that the point mass approximation
for stars was neglecting a number of dynamically significant processes
in clusters. Allowing stars to have radii, and having those radii
change with the evolution of the star, was an important next addition
to stellar dynamics simulations \citep{HI85, A96, PMHM01, HTAP01}. Finite
stellar radii are most important for two aspects of these
simulations. First, binary stars can undergo mass transfer as one
member of the system fills its Roche lobe, either through evolution of
the star or dynamical modification of the orbital parameters of the
system.  Altering the parameters of binary systems will change how the
binaries affect the evolution of the cluster. In the extreme case, the
two components of the binary system can merge. Secondly, stars with
finite radii can collide with other stars. The low velocity stellar
collisions that occur in star clusters produce blue stragglers
\citep{SLBDRS97} and could produce other non-standard stellar
populations. These populations in turn can change the local dynamical
evolution in the cluster.

Until recently, all work on cluster modelling assumed that all stars
began their lives on the main sequence. However, low mass stars make
up the bulk of stars in a cluster for any reasonable initial mass
function. Low mass stars also have significant pre-main sequence
lifetimes. For some young open clusters, most of the stars are still
on the pre-main sequence. These young stars have radii which can be up
to $\sim$ 10 times larger than their main sequence radii
\citep{SDF00}. Therefore, some binaries could have undergone an
episode of mass transfer that is not taken into account. Also, larger
stars are more likely to have experienced a collision; those collision
products will have been missed in previous simulations. Dynamical
simulations which include a pre-main sequence phase show that there
can be notable effects on the overall cluster evolution (Wiersma,
Sills \& Portegies Zwart, in preparation). However, in order to
properly include the pre-main sequence phase in stellar dynamics
calculations, we need to understand collisions with pre-main sequence
stars, and the properties of the collision products. Currently no such
calculations are available in the literature.

In this paper, we present smoothed particle hydrodynamic (SPH)
simulations of direct collisions between pre-main sequence stars and a
variety of other stellar populations (main sequence, giant branch,
compact objects). We study collisions at a variety of impact
parameters and velocities, and determine the structure and likely
evolution of the collision products. In section 2, we outline our
simulation method and initial conditions. We present the results in
section 3, and discuss their implications for stellar dynamics in
section 4.

\section{Method}

The collisions studied in this paper were simulated using the
smooth particle hydrodynamics (SPH) method
\citep{B90,M92}, where the SPH particles represent the stellar gas. The 
three dimensional code we used is described in detail in
\citet{SADB02}, and is the parallel version of the code described in
\citet*{BBP95}. It uses a tree to solve for the gravitational forces
and to find the nearest neighbours \citep{B90}. We use the standard
form of artificial viscosity with $\alpha = 1$ and $\beta = 2.5$
\citep{M92}, and an adiabatic equation of state. The thermodynamic
quantities are evolved by following the change in internal
energy. Both the smoothing length and the numbers of neighbours can
change in time and space. The smoothing length is varied to keep the
number of neighbours approximately constant ($\sim 50$).

\begin{figure}
\epsscale{1.0}
\plotone{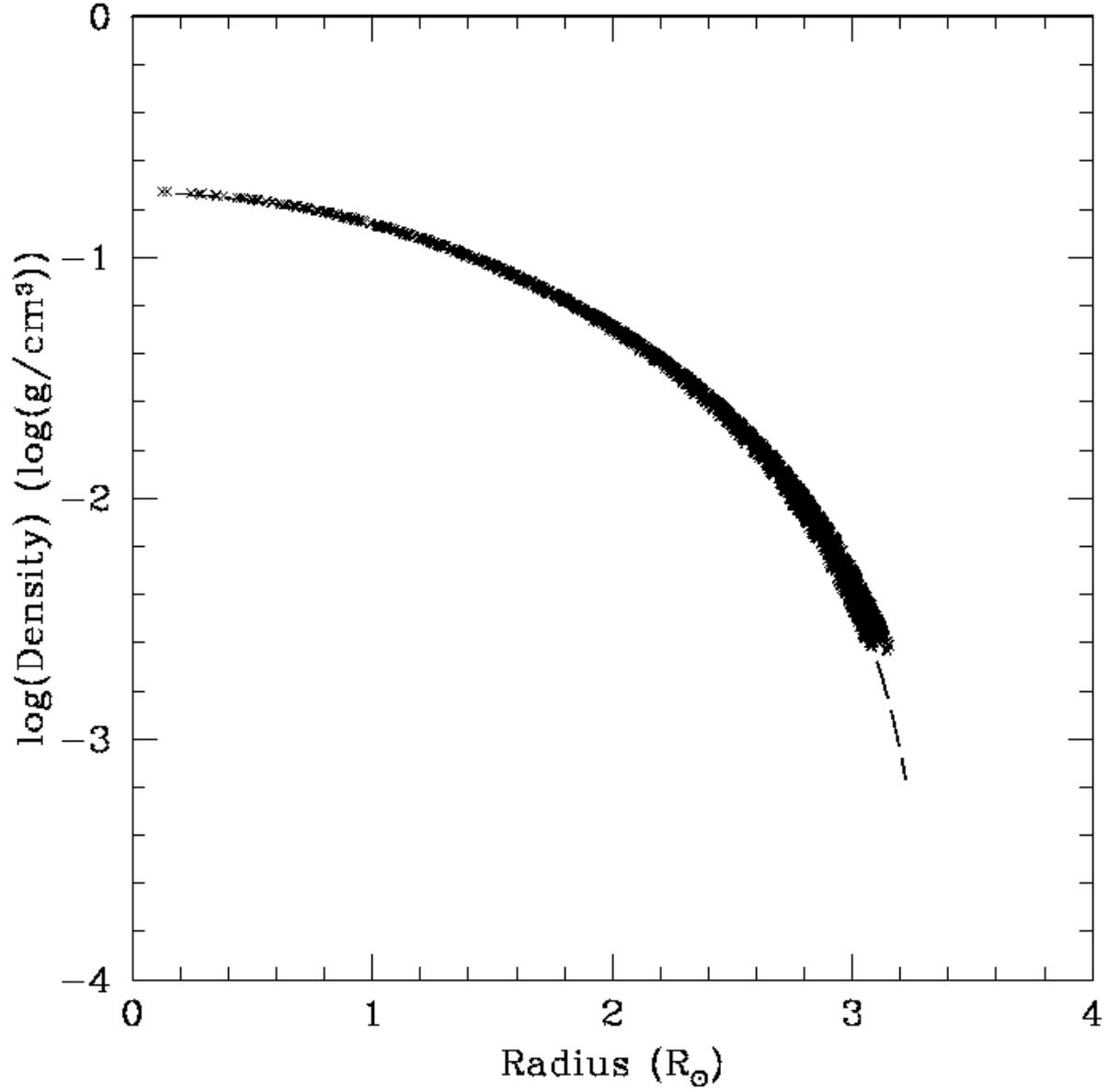}
\caption{\footnotesize Logarithm of the density versus radius for the 
pre-main sequence star used in all simulations.  The dashed line
is the stellar model from YREC while the x's are the SPH
representation of this star.\label{PMSdensity}}
\end{figure}

Collisions involving five different types of stars were examined in
this paper.  Specifically, we simulated interactions between pre-main
sequence (PMS) stars and white dwarfs (WD), giant branch (GB),
zero age main sequence (ZAMS), turnoff main sequence (TAMS),
and other pre-main sequence stars. The density and composition
profiles of the stellar models were calculated using the Yale
Rotational Evolution Code (YREC, \citet{GDKP92}). The stars were
composed of $\sim$ 10,000 SPH particles except for the giant which
required $\sim$ 50,000 particles in order to properly model its
steeper density gradient.  The giant also contained a point mass core
to model the very high density in this region.  The white dwarf was
modeled as a point mass due to its very small radius and large density
compared to the other stars studied. The SPH particles were equally spaced
throughout the star, and their masses were varied to produce the
required density profile.  The stars were all run alone in the SPH
code in order to allow them to relax before being collided. Figure
\ref{PMSdensity} shows the density profile of the SPH particles for
the pre-main sequence star used in all collisions, as well as the YREC
model (dashed line). This pre-main sequence star is barely off the
deuterium-burning birthline, and has an age of $\sim$ 100 000
years. Pre-main sequence stars actually spend little time in this part
of the HR diagram. Our 0.8 \msun model spends 70 Myr contracting to
the main sequence, and only the first 10 Myr of that time is on the
Hayashi track. The rest of the time is spent traveling horizontally
across the HR diagram. The structure of the pre-main sequence star
during this horizontal phase is very similar to that of a main
sequence star. Therefore, in order to expand our understanding of
stellar collision products with very different properties, we chose an
extremely young PMS star for our collision simulations.

Table \ref{parentstars} gives the parameters of all the parent stars
studied in this paper.  The number of SPH particles used in each model
is given in units of 1,000 particles.  All stars are 0.8 \msun~, except the
white dwarf which is 0.6 \msun~.
The stars have solar metallicity, Z=0.0188, and an initial helium abundance
of Y=0.27. The giant branch star was evolved to about halfway up the giant
branch. This giant was chosen as a representative average giant that
might collide with another object -- the cross section for collision
increases as the giant's radius increases, but its lifetime at that
radius decreases.

\begin{deluxetable}{cccccc}
\tabletypesize{\scriptsize}
\tablecaption{Table of Star Parameters\label{parentstars}}
\tablewidth{0pt}
\tablehead{
\colhead{Star Name} & \colhead{Star Type} & \colhead{Mass}  & \colhead{Radius} & \colhead{Number of SPH particles} \\
 &  & \colhead{\msun}  & \colhead {\rsun} & \colhead {1000's}
}
\startdata
PMS1 & Pre-main Sequence & 0.8  & 3.4 & 10 \\
PMS2 & Pre-main Sequence & 0.8  & 3.4 & 50 \\
PMS3 & Pre-main Sequence & 0.8  & 3.4 & 100 \\
ZAMS & Zero Age Main Sequence & 0.8  & 0.72 & 10 \\
TAMS & Terminal Age Main Sequence & 0.8 &  1.03 & 10 \\
GB & Giant Branch & 0.8 & 7.01 & 50 \\
WD & White Dwarf & 0.6 & 0.001 & 0 \\
\enddata
\tablecomments{The giant branch star contains a point mass core of approximately 0.22 \msun~ and 0.028 \rsun.  The SPH particles model the remaining mass.  Also, the white dwarf is modeled as a point mass.}
\end{deluxetable}

It is, of course, unreasonable to expect stars of the same mass to be
simultaneously on the pre-main sequence and on the giant branch in the
same cluster. However, in this paper we are interested in the gross
properties of collisions involving pre-main sequence stars. At that
level, the results of collisions between stars of different masses can
be extrapolated from the results presented here. 

\begin{figure}
\epsscale{1.0}
\plotone{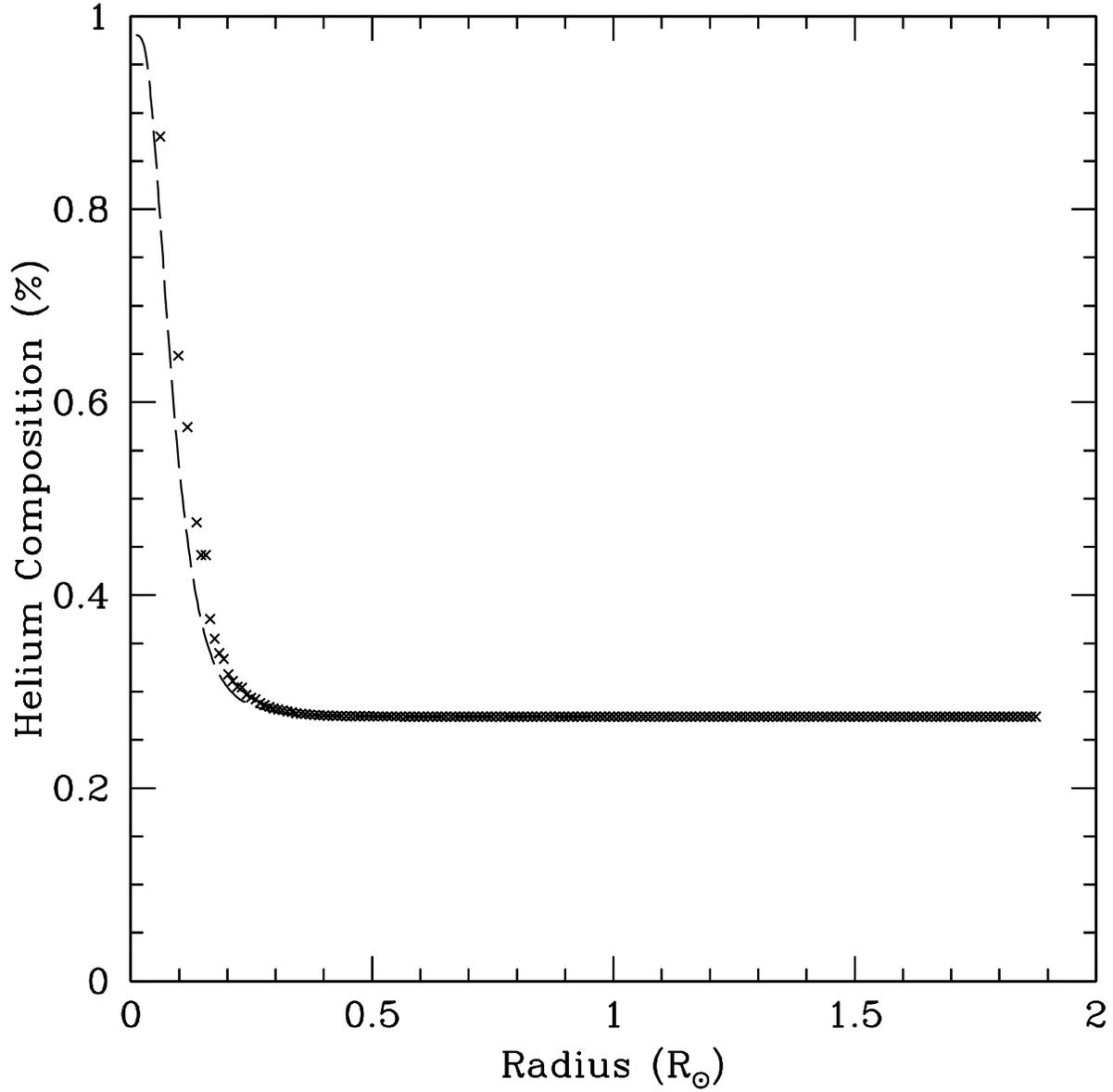}
\caption{\footnotesize Helium composition as a function of radius for the 
TAMS star used in simulations.  The dashed line is the stellar
model from YREC and the x's are the SPH representation of this star.\label{TAMShe}}
\end{figure}

The helium composition profile for the terminal age main sequence
(TAMS) star is displayed in figure \ref{TAMShe}. The helium
composition profiles of the pre-main sequence, zero age main sequence,
and giant branch stars are flat. This is true for the GB
because all of the helium enriched fluid is within the point mass
core, so the SPH particles in the rest of the star all have essentially the
same content.

In this paper, we consider all five of the
stars as impactors to the pre-main sequence star.  The frame of
reference used for simulations was such that the stars' center of mass
initially resided at the origin.  An initial separation of 30
\rsun~ was used to ensure that in all initial setups the stars were
separated by at least 4 times the radius of the larger star.  This
allowed tidal effects to be ignored in the original alignment.  The
stars were set up on a zero-energy orbit initially under the
assumption that the stars were point masses. The relative velocities
at infinity of the stars ranged from 10km/s to 100km/s, spanning the
range of typical velocities found in collisions in clusters.  The
lower bound, 10 km/s, is the typical velocity dispersion in globular
clusters. The upper bound, 100 km/s, is a representative upper bound
on the velocity a binary system could provide in a resonance encounter
which resulted in a physical collisions. The velocity dispersion in
open clusters is smaller than our lower bound (typically a few
km/s). For collisions with low velocities at infinity, the actual
impact speed is dominated by the escape velocity at the surface of the
two stars. We expect our results for 10 km/s to be a reasonable
approximation to all low velocity collisions. Initially, the stars
were nonrotating, so any angular momentum after the collision was
purely a result of the initial orbit.  The orbital plane was chosen to
be the z=0 plane.

The impact parameter $r_p$ was varied from 0, for a head-on collision,
to 1 (in units of R$_1$ + R$_2$, the radii of the two stars), for a
grazing collision, in order to gain an understanding of the results for
collisions ranging over the whole radius of the star.  Collisions with
an impact parameter greater than 1 could result in a merged daughter
star, due to the fact that tides are raised on the stars as they
approach each other, dissipating energy and possibly causing contact
as they pass.  This case was not considered, however, due to its much
larger computational requirements.  If the first pass of the stars
does not dissipate enough energy then the stars will take a substantial
amount of time to merge because they can convert more kinetic
energy back into potential energy and therefore achieve larger separations.  Obviously, if
the impact parameter is larger than the combined radius then very
little energy will be dissipated on the first pass.  Also, some
impactors were not considered at an impact parameter of 1 if their
companions at an impact parameter of 0.5 took a long time to merge,
also due to the large computational requirements.

\section{Results}
Table \ref{results} summarizes the results of the simulations
performed.  In column 1, we give a letter which names each run.
Column 2 indicates the impactor of the PMS star used for the run.
Column 3 gives the impact parameter, in units of the total radii of
the two stars, while column 4 gives the initial velocity at infinity
of the pair.  The remaining data provides information about the
daughter star produced.  Column 5 indicates if the stars merged before
the simulation was ended. Columns 6-8 display the daughter stars'
masses in units of \msun, final angular momentum around the z-axis in
units of $g~ cm^2s^{-1}$, and the ratio of rotational kinetic energy
to absolute potential energy respectively.  The final column indicates
the percent change in total energy throughout the course of the
simulation.

\begin{deluxetable}{ccccccccc}
\tabletypesize{\scriptsize}
\tablecaption{Table of Run Results\label{results}}
\tablewidth{0pt}
\tablehead{
\colhead{Case} & \colhead{Impactor} & \colhead{($\frac{r_{p}}{R_{1}+R_{2}}$)} & \colhead{(v$_{inf}$)} & \colhead{Merged} & \colhead{M$_{final}$} & \colhead{J$_{final}$} & \colhead{$\frac{T}{|W|}$} & \colhead{Energy Conservation} \\
& & & \colhead{km/s} & & \colhead{\msun} & \colhead{g$\frac{cm^2}{s}$} & & 
}
\startdata 
A & PMS1 & 0.0 & 10 & Y & 1.55 & 6.77E48 & 1.47E-06 & 5.31E-3\%\\
B & PMS1 & 0.5 & 10 & Y & 1.60 & 7.92E51 & 1.49E-2 & 6.69E-1\%\\
C & PMS1 & 1.0 & 10 & Y & 1.60 & 1.12E52 & 3.53E-2 & 8.44E-1\%\\
D & PMS1 & 0.0 & 40 & Y & 1.55 & 4.23E48 & 1.65E-6 & 3.16E-2\%\\
E & PMS1 & 0.5 & 40 & Y & 1.59 & 7.92E51 & 1.31E-2 & 5.94E-1\%\\
F & PMS1 & 1.0 & 40 & Y & 1.60 & 1.12E52 & 4.23E-2 & 7.89E-1\%\\
G & PMS1 & 0.0 & 100 & Y & 1.54 & 3.97E48 & 2.12E-6 & 9.31E-2\%\\
H & PMS1 & 0.5 & 100 & Y & 1.58 & 7.98E51 & 1.50E-2 & 6.18E-1\%\\
I & PMS1 & 1.0 & 100 & N & - & - & - & 7.79E-1\%\\
J & ZAMS & 0.0 & 100 & Y & 1.47 & 7.81e+49 & -1.09E-5 & 8.28E-1\%\\
K & ZAMS & 0.5 & 100 & Y & 1.48 & 2.02e+51 & -7.59E-4 & 8.34E-2\%\\
L & TAMS & 0.0 & 10 & Y & 1.45 & 3.36E50 & 1.29E-5 & 4.31\%\\
M & TAMS & 0.5 & 10 & Y & 1.51 & 5.97E51 & 1.95E-3 & 4.15E-1\%\\
N & TAMS & 0.0 & 40 & Y & 1.44 & 2.53E50 & 9.86E-6 & 1.47\%\\
O & TAMS & 0.5 & 40 & Y & 1.51 & 5.86E51 & 1.99E-3 & 3.29E-1\%\\
P & TAMS & 0.0 & 100 & Y & 1.43 & 3.18E50 & 8.58E-6 & 8.62E-1\%\\
Q & TAMS & 0.5 & 100 & Y & 1.49 & 5.61E51 & 2.00E-3 & 1.92\%\\
R & GB & 0.0 & 10 & Y & 1.40 & 3.28E49 & 1.57E-5 & 1.00\%\\
S & GB & 0.5 & 10 & Y & 1.54 & 8.95E51 & 1.59E-2  & 4.60E-1\%\\
T & GB & 0.0 & 40 & Y & 1.40 & 2.23E49 & 1.47E-5 & 4.13E-1\%\\
U & GB & 0.5 & 40 & Y & 1.53 & 8.89E51 & 1.67E-2 & 5.40E-1\%\\
V & GB & 0.0 & 100 & Y & 1.38 & 1.76E49 & 1.60E-5 & 7.39E-2\%\\
W & WD & 0.0 & 10 & Y & 1.16 & 6.65E49 & 1.69E-5 & 9.10\%\\
X & WD & 0.5 & 10 & Y & 1.29 & 3.47E51 & 9.67E-3 & 12.9\%\\
Y & WD & 1.0 & 10 & Y & 1.28 & 4.72E51 & 1.38E-2 & 13.6\%\\
Z & WD & 0.0 & 40 & Y & 1.16 & 3.82E49 & 1.66E-5 & 14.1\%\\
AA & WD & 0.5 & 40 & Y & 1.29 & 3.52E51 & 1.01E-2 & 12.8\%\\
AB & WD & 1.0 & 40 & Y & 1.30 & 5.07E51 & 1.56E-2 & 10.1\%\\
AC & WD & 0.0 & 100 & Y & 1.15 & 5.56E49 & 1.70E-5 & 11.8\%\\
AD & WD & 0.5 & 100 & Y & 1.29 & 3.66E51 & 1.06E-2 & 12.7\%\\
AE & WD & 1.0 & 100 & Y & 1.27 & 4.96E51 & 1.58E-2 & 10.4\%\\
\enddata
\end{deluxetable}

\begin{figure}
\epsscale{1.0} 
\plotone{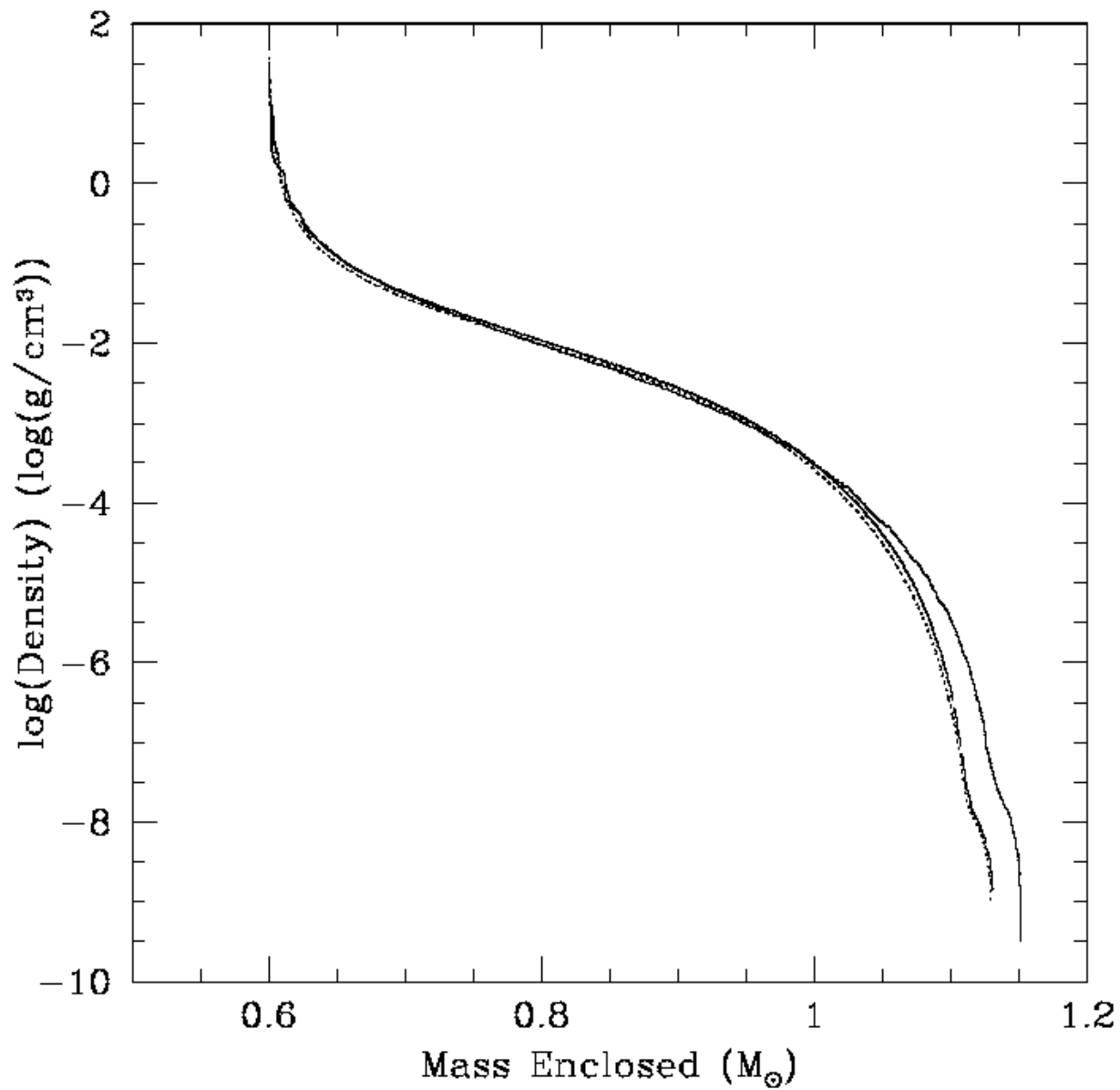}
\caption{\footnotesize Logarithm of the density versus mass enclosed by an isodensity surface
for the product of run AC with three different resolutions.  The solid line
is the product of a pre-main sequence star with $\sim$ 10 000 particles, the long
dashed line had $\sim$ 50 000 particles in the parent PMS star, and the
dotted line is the result of the same run with a $\sim$ 100 000
particles PMS star. \label{resolution}}
\end{figure}

\subsection{Numerical Resolution}

Case AC, a head-on collision between a pre-main sequence star and a
white dwarf, was run with three pre-main sequence stars of varying
number of particles to test the effects of resolution on the results
we obtained.  Along with the 10 000 particle case described in the
previous section, two additional higher resolutions, 50 000 and 100
000 particles, were run.  Figure \ref{resolution} shows a plot of the
final density profiles for these three runs.  The solid line shows the
10 000 particle case, the long dashed line is the result of the 50 000
particle case, and the dotted line is the result of the 100 000
particle case.  This plot shows that the resolution of the pre-main
sequence star did not significantly alter the final product,
especially in the central region of the daughter star.  There are no
noticeable differences in the inner 70\% by mass of these three
density profiles.  Slight disagreement is visible in the outer regions
of the star. However, this region is the least settled and will be
subject to further relaxation. The density profile of the lowest
resolution at future times shows that the outer region of the star is
indeed slowly decreasing in density as the star continues to relax,
while the density at the core is not changing at all.  This, as can be
seen in the plot, would bring it into closer agreement with the other
resolutions. It is important to note that although there are
noticeable differences between the 10 000 and 50 000 particle
resolutions, the 50 000 and 100 000 resolutions are essentially
indistinguishable.  This indicates that the resolution we used in our
simulations (10 000) was an adequate approximation since the daughter
stars converge, as the number of particles is increased, to a model
very similar to the one we achieved.  Increasing the resolution simply
further refined the results, not drastically altered them.  As a
result of this test, we concluded that while a resolution of 50 000
particles per parent star is optimal for an accurate description of
the structure of the collision product (in agreement with
\citet{SADB02}), a lower resolution of 10 000 particles is sufficient
to determine the product's general properties.

\begin{figure}
\epsscale{1.0}
\plotone{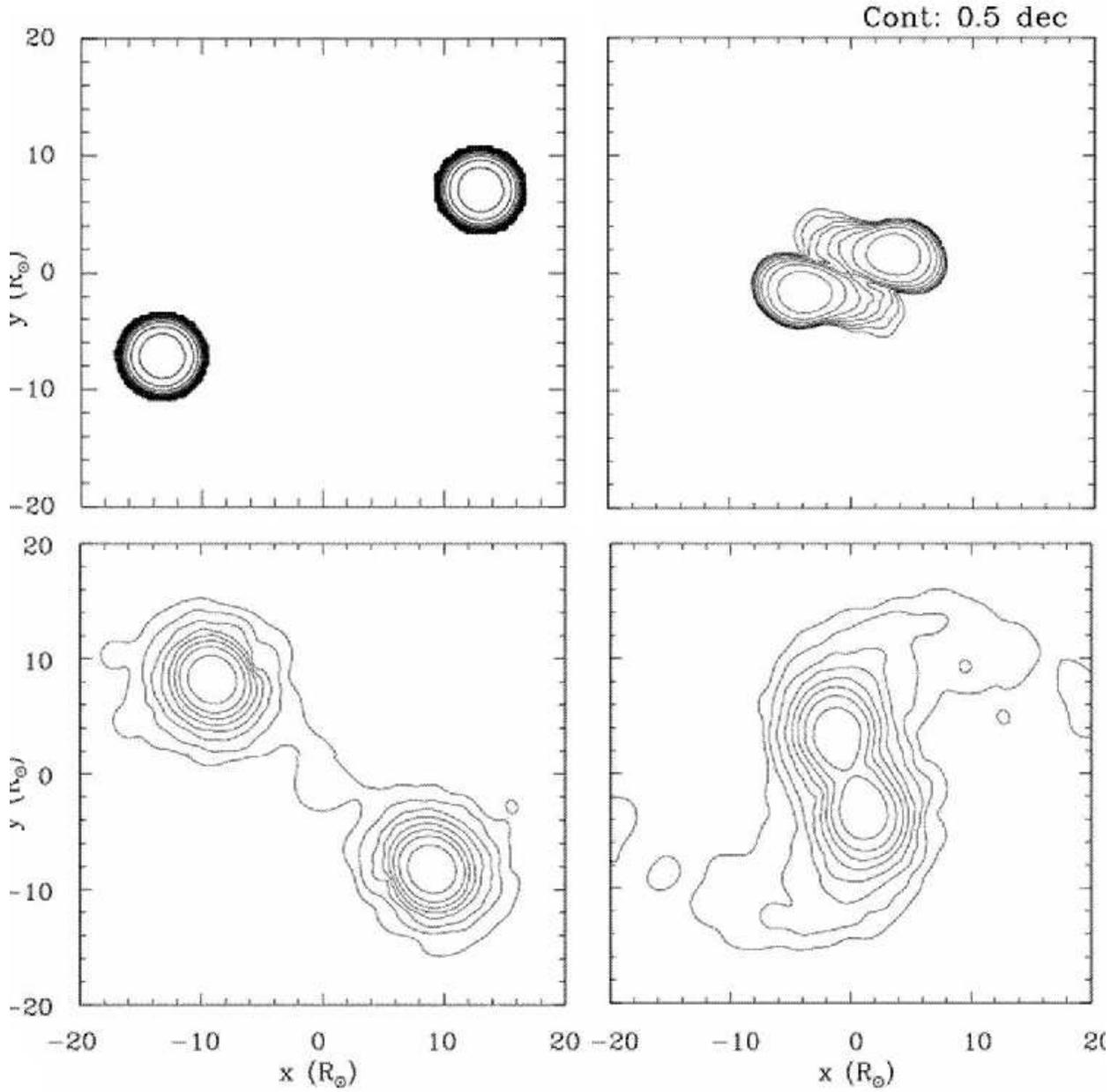}
\caption{\footnotesize Density contours in the orbital plane spaced logarithmically from several stages of collision C.  There are eight contours equally spaced over four decades down from the maximum density. The top left plot is at t=0 hours, the top right is at 39 hours, the bottom left is at 542 hours, and the bottom right is at 690 hours.\label{contours1}}
\end{figure}

\begin{figure}
\epsscale{1.0}
\plotone{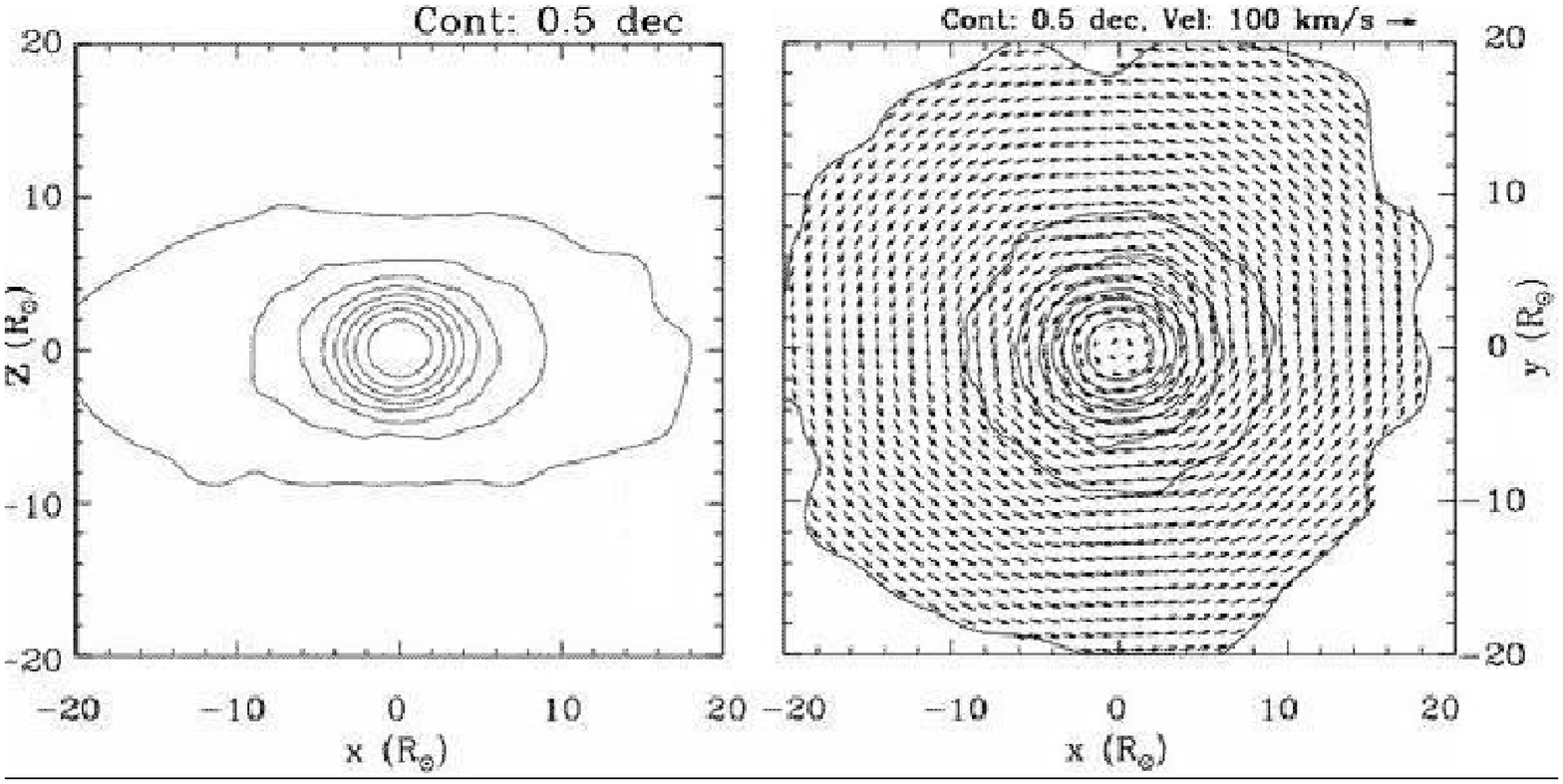}
\caption{\footnotesize Density contours in the xz plane, spaced logarithmically, and velocity field in the orbital plane for the final product of collision C. Both plots are at the end time of the simulation: t=1767 hours after the simulation began.\label{contours3}}
\end{figure}

\subsection{A Representative Case}

As a representative case, we will describe run C in detail.  This is a
collision between two PMS stars with an impact parameter of 1, and a
relative velocity at infinity of 10 km/s.  This collision took approximately
250 CPU hours to simulate.  Figures
\ref{contours1}-\ref{contours3} show the dynamical evolution through
density contours at different stages of the collision.  Despite the
fact that the parameters predict that the collision would be just
grazing, tidal effects caused the stars to be misshapen on approach and
hence significant contact is made during the first pass. The stars go
through 5 periastron passages before merging at a time of 730
hours. The collision product is rotating quickly, and is flattened by
rotation, as seen in the x-z plane shown in figure
\ref{contours3}b. There is an extended halo of material around the
star, but it is not sufficiently flattened to be called a disk. The
bulk of the material is within a radius of about 7 \rsun. This is
significantly larger than the radius of a typical 1.6 \msun  main
sequence star, but is comparable to a radius of a 1.6 \msun  pre-main
sequence star on the Hayashi track. We expect that when this collision
product contracts to the main sequence, it will be reasonably
indistinguishable from a normal main sequence star. 

\begin{figure}
\epsscale{1.0}
\plotone{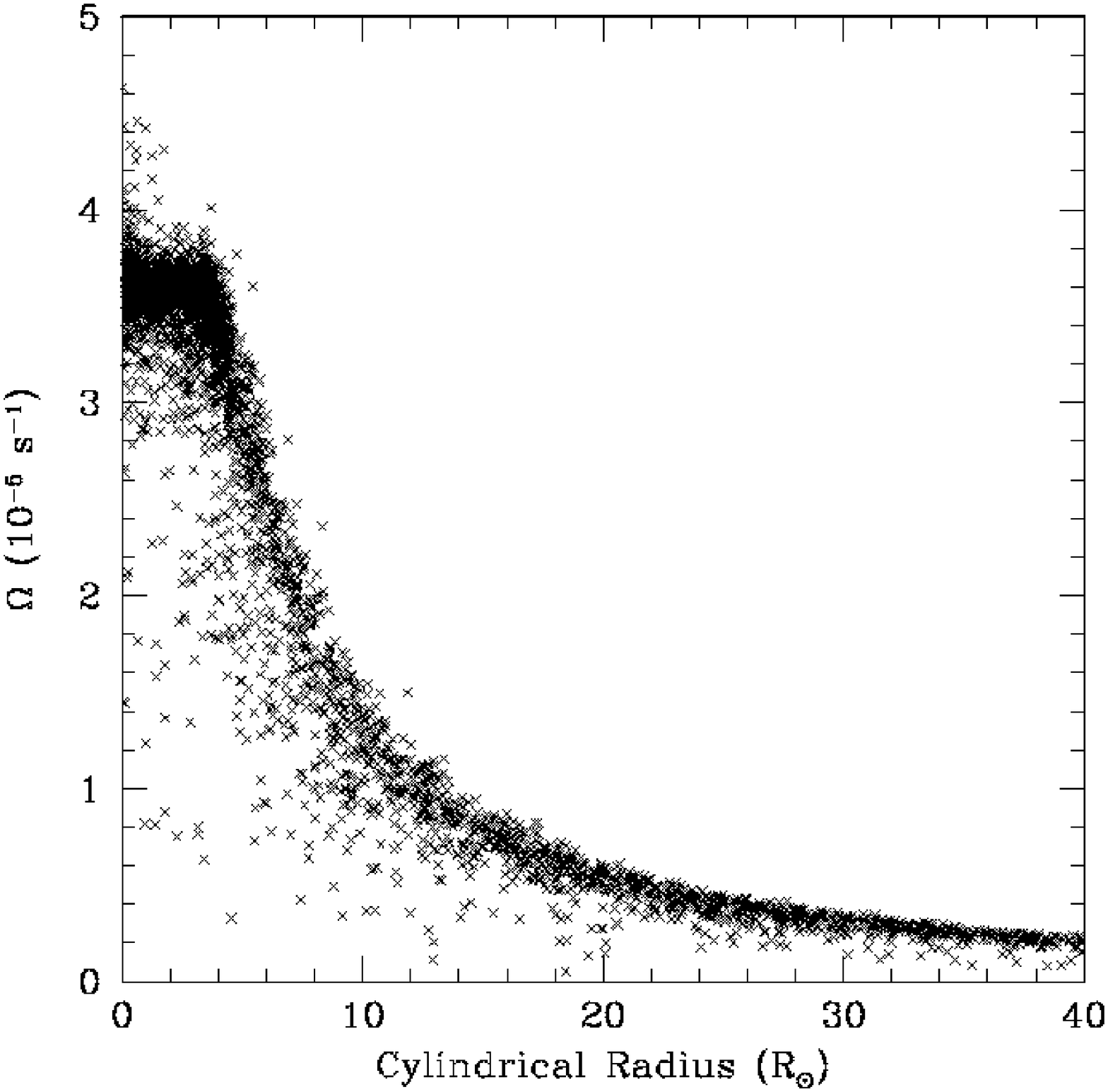}
\caption{\footnotesize Angular velocity as a function of cylindrical radius for the final product of collision C. \label{angvel}}
\end{figure}

Figure \ref{angvel} shows the angular velocity as a function of
cylindrical radius, the distance in the orbital plane to the rotation
axis, of the daughter star of run C.  This angular velocity data is only shown
for particles within 2 smoothing lengths of the equatorial plane.
This figure shows that the central region of the star is rotating
at a constant rate.  At a radius of about 5\rsun~ the rotational
velocity begins to drop and reaches a minimum at the edge of the star.

\begin{figure}
\epsscale{1.0}
\plotone{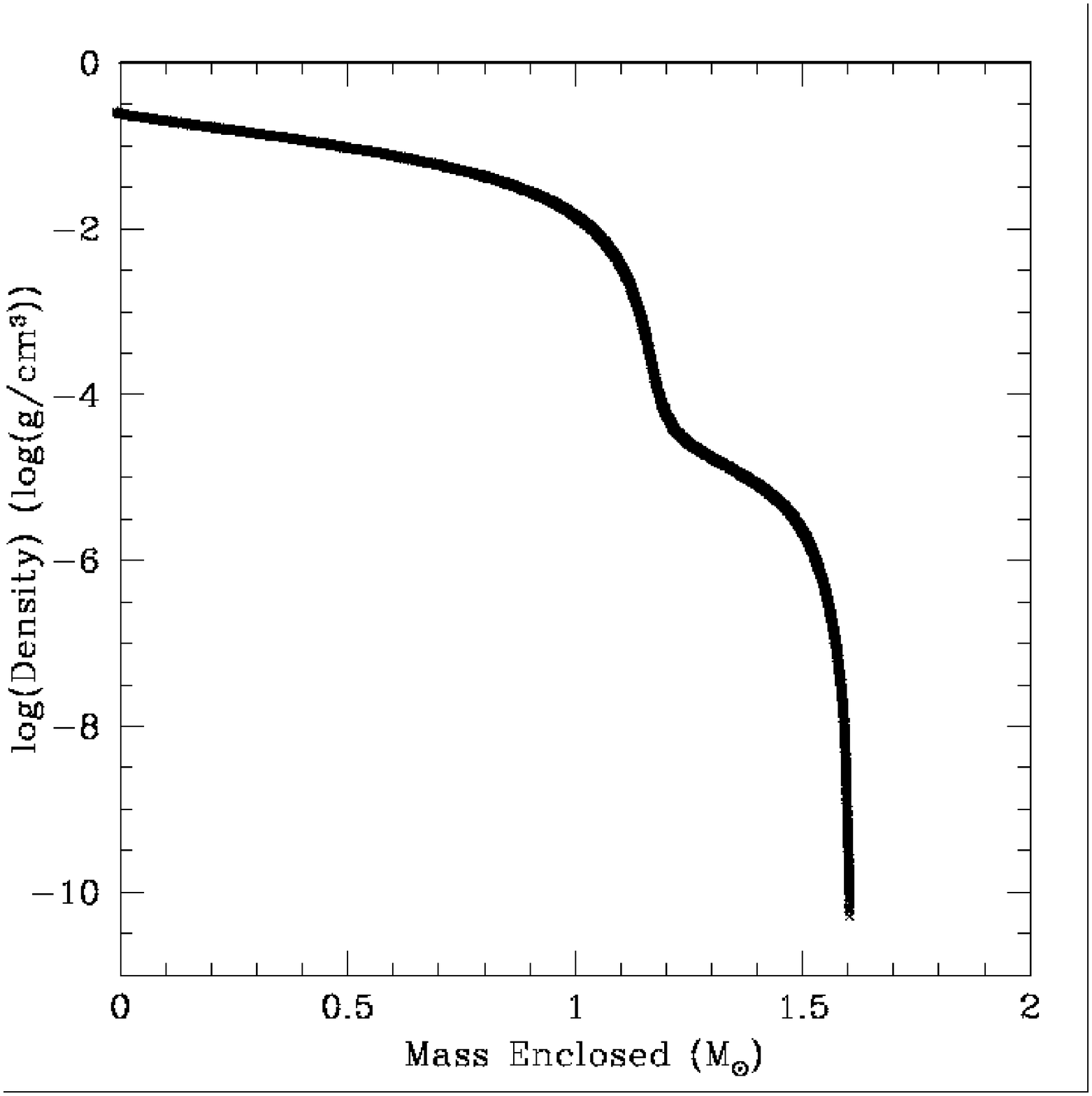}
\caption{\footnotesize Density as a function of mass enclosed by isodensity surfaces for the final product of collision C. \label{densityC}}
\end{figure}

\begin{figure}
\epsscale{1.0}
\plotone{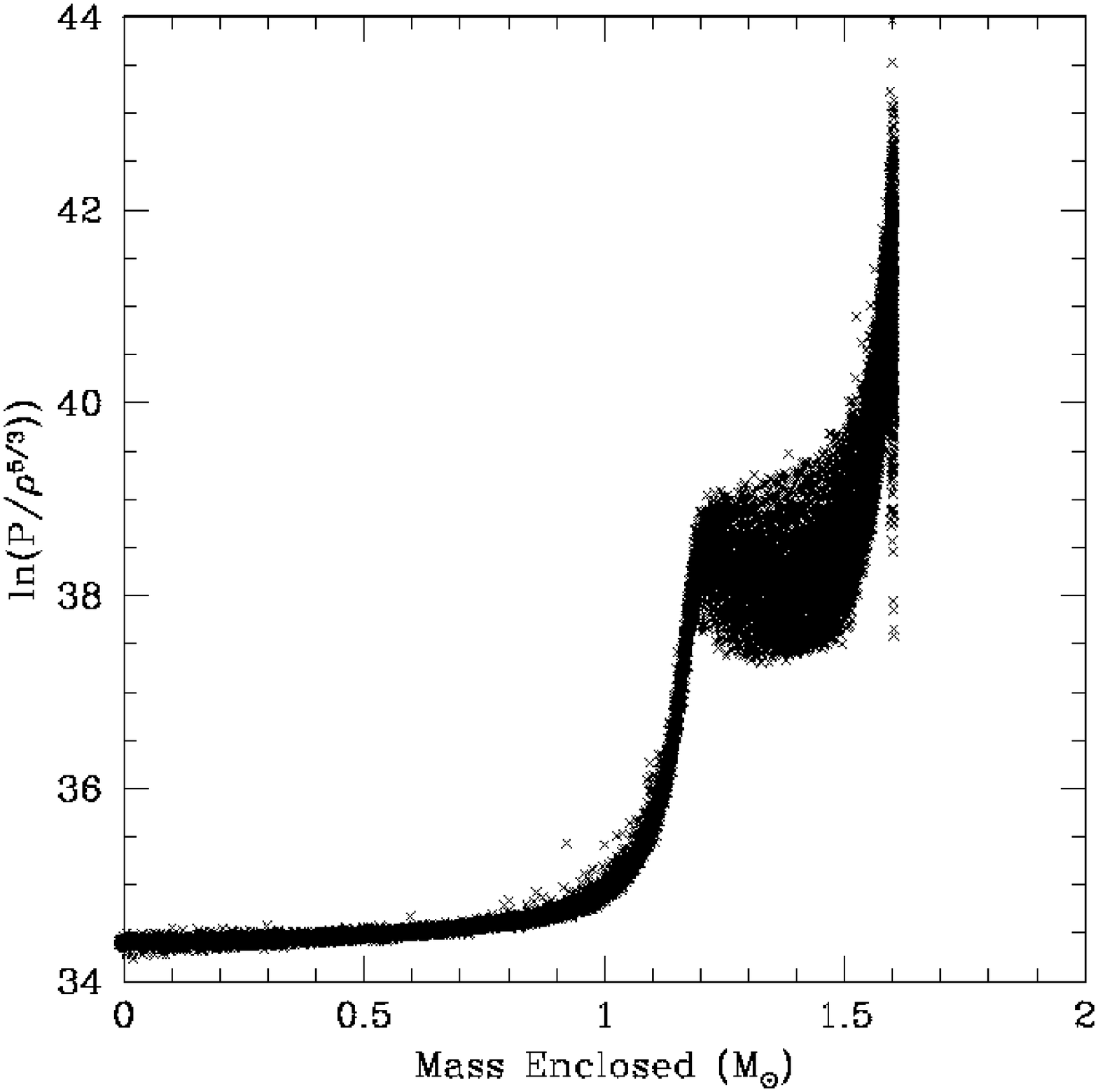}
\caption{\footnotesize Entropy as a function of mass enclosed by isodensity surfaces for the final product of collision C. Pressure and density are measured in cgs units.  \label{entropyC}}
\end{figure}

Figure \ref{densityC} displays the density profile and figure
\ref{entropyC} shows the entropy profile of the resulting star.
Ideally, the entropy profile will be such that as you look at larger
enclosed masses, the entropy always increases (ignoring the extreme
edge of the star).  This would imply that the inner majority of the
star is in hydrostatic equilibrium \citep{LRS96}.  Initially after the
collision this is not the case; the entropy versus mass plots
are very disordered.  As the star relaxes, the inner region of the
star begins to develop a well defined entropy curve. The region that is
in hydrostatic equilibrium grows from the centre outward until this
well defined curve extends essentially all the way to the edge of the
star.  However, we do not continue our simulations until the entire star
is in hydrostatic equilibrium -- the computational requirements are
too large and we would be introducing unacceptably large errors in
the total energy if we continued the simulation for the many more
dynamical times required to bring the entire star into
equilibrium. These simulations are sufficient to describe the
behaviour of the inner 70\% by mass of the daughter star.

\begin{figure}
\epsscale{1.0}
\plotone{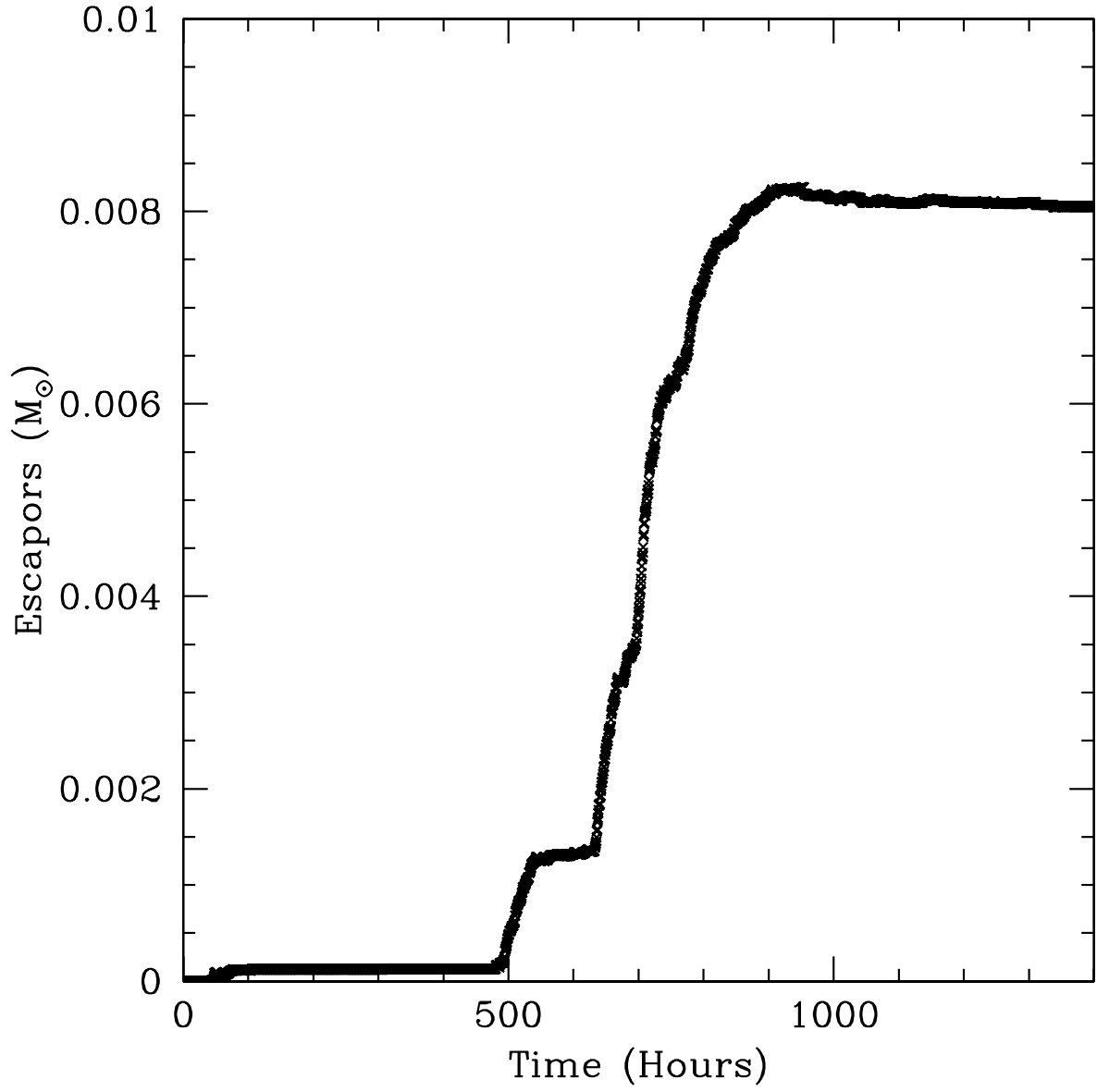}
\caption{\footnotesize Escaped mass as a function of time for collision C. \label{escape}}
\end{figure}

In figure \ref{escape} the escaping mass as a function of time for run C is
displayed.  Each jump in the escaping mass represents a point of closest
approach of the two stars.  By the 4th pass the two stars are
essentially merged into one. In this collision, only 0.5\% of the
total mass is lost from the system.

\begin{figure}
\epsscale{1.0}
\plotone{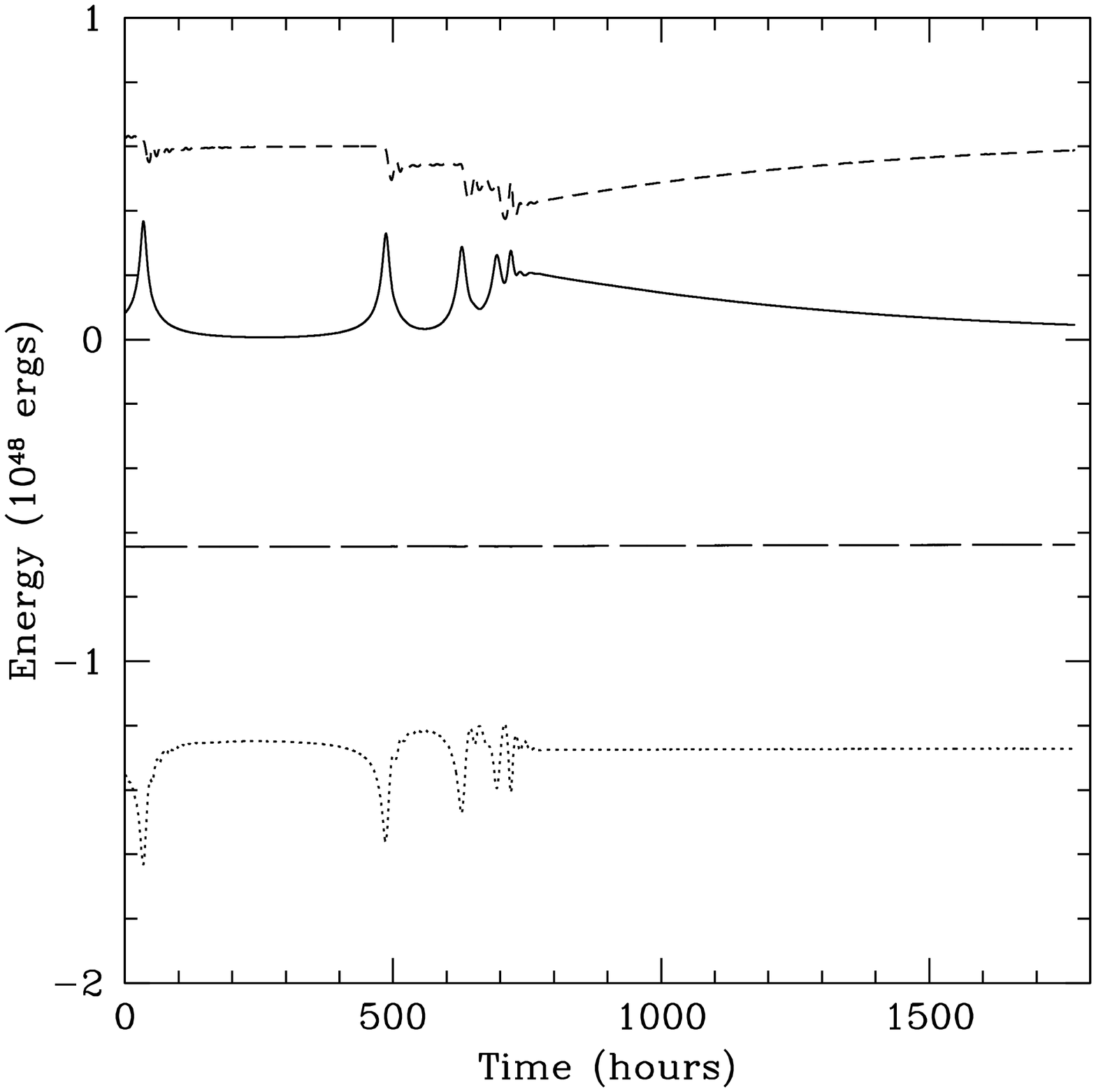}
\caption{\footnotesize Energies as a function of time for collision C.  The 
doted line is the potential energy, the solid line is the kinetic energy,
the long dashed line is the total energy, and the short dash line is
the internal energy.\label{energy}}
\end{figure}

The energy plots for this collision followed a pattern typical to all
cases studied.  The energy as a function of time can be seen in figure
\ref{energy}.  Total energy was conserved in this run to better than
1\%.  The internal, potential and kinetic energies fluctuated as the
collision proceeded.  The dips in potential energy and internal
energy, which were mirrored by jumps in kinetic energy, correspond to
points of closest approach of the two orbiting stars.  As the stars
moved towards their first closest approach point, potential energy was
converted to kinetic energy, as the two stars sped up. At the moment
of impact, energy is converted to internal energy, as a result of shock
heating, at the expense of kinetic energy which is lost during the
collision.  As the two stars pass each other and begin to separate,
the reverse process occurs with the potential increasing as the
kinetic decreases.  Each point of closest approach leads to another
cycle of this process, with each time the minimum potential energy
increasing and the maximum kinetic energy decreasing as the stars
spiral into each other.  After the third maximum in potential energy,
the peaks in potential energy begin to decrease indicating that the stars are
orbiting in a common envelope of material.  Once the spikes have disappeared,
the star has finally completed merging.  Over the long run, as the star
settles down, the kinetic energy was slowly lost to internal energy.

\subsection{Results of the Other Simulations}

The other runs give results similar to the run described above, with
obvious differences dictated by the differences in impactor.

\begin{figure}
\epsscale{1.0}
\plotone{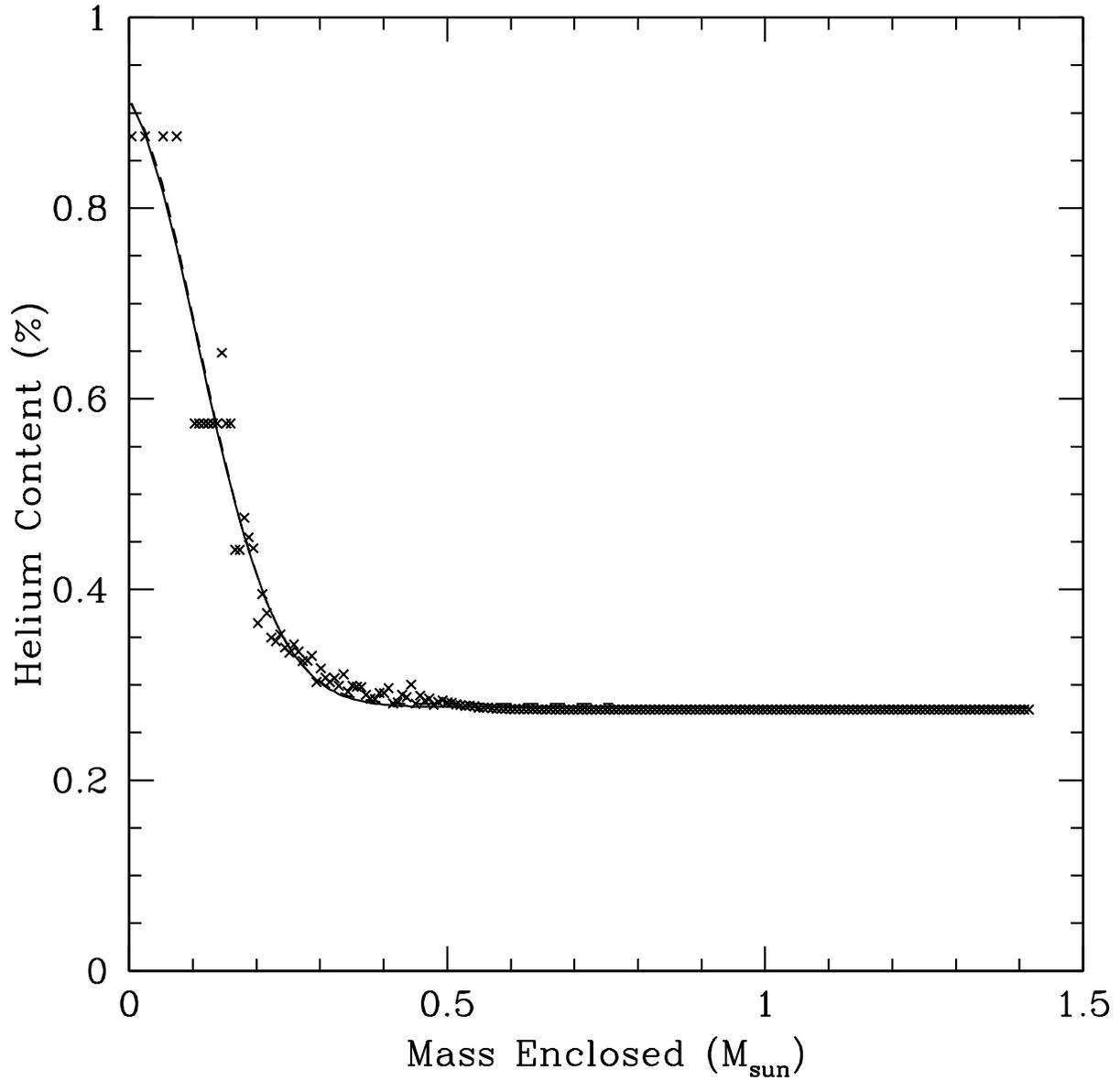}
\caption{\footnotesize The helium composition as a function of enclosed mass for the collision product of run P. The x's are the average helium composition values for bins of specific radii and the solid line is a curve fitted to this data.  The dashed line is the helium composition profile for the parent TAMS star. \label{composition}}
\end{figure}

The only collisions which produced a star with a non-constant helium
composition profile were those between the TAMS and PMS star, due to the
fact that the TAMS star was the only parent star with a varying helium
composition profile.  Figure \ref{composition} shows a plot of the helium
composition for the result of run P.  This was calculated by
averaging the helium composition over annuli centered at the center
of the daughter star.  This plot shows that the overall shape of the
profile remains approximately the same as the original TAMS star.

\begin{figure}
\epsscale{1.0}
\plotone{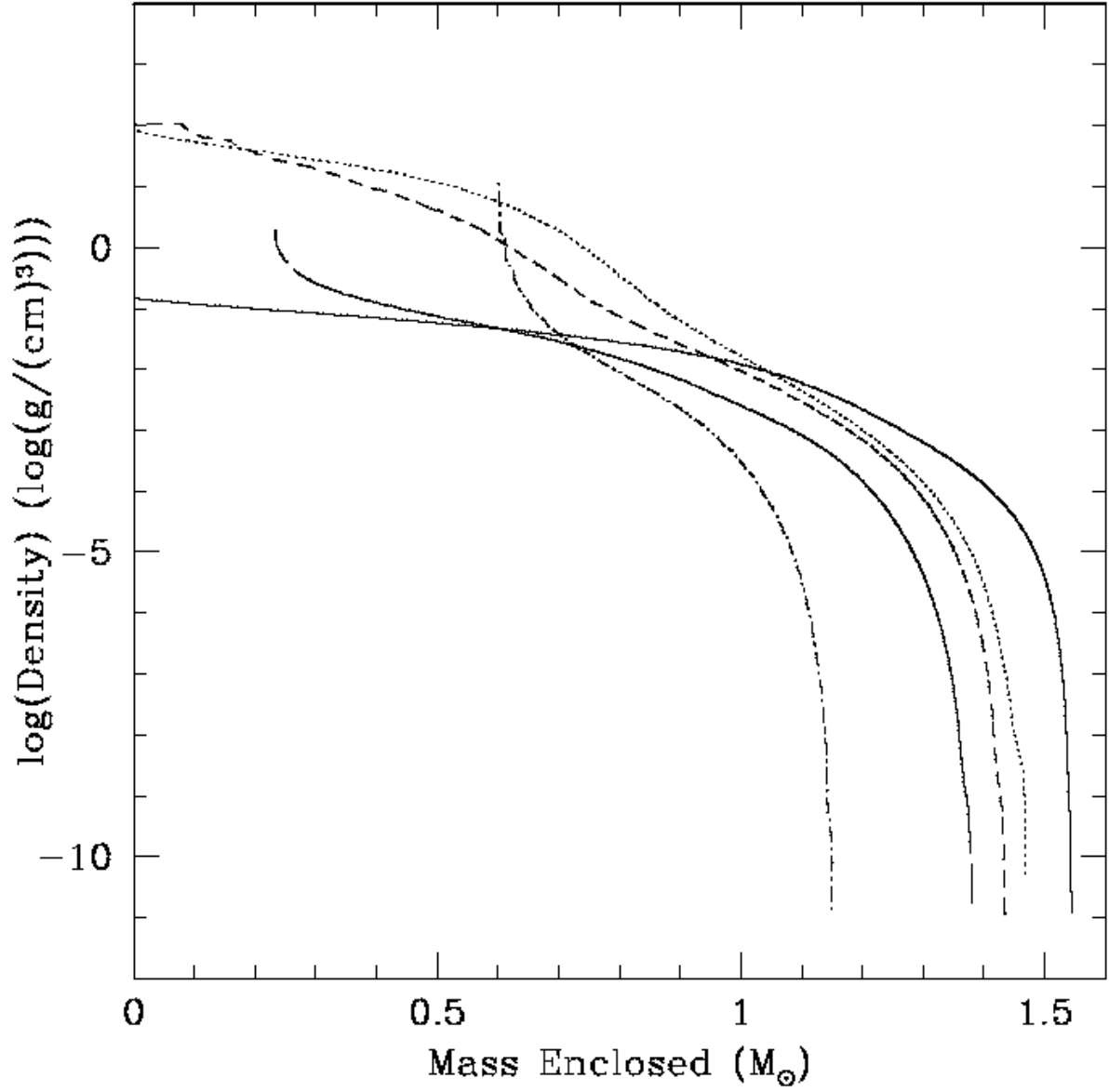}
\caption{\footnotesize Density as a function of mass enclosed for collisions G (PMS+PMS, solid line), J (PMS+ZAMS, dotted), P (PMS+TAMS, short dash), V (PMS+GB, long dash), and AC (PMS+WD, dash-dot). \label{densityprofiles}}
\end{figure}

Figure \ref{densityprofiles} shows the density profiles of collision
products of head-on collisions (with v=100km/s) between pre-main
sequence stars and other objects. In this figure, only the SPH
particles have been included. Some simulations include a point mass
(e.g. the white dwarf and the core of the giant). Those collision
products have density profiles which start at masses greater than
zero, with the point mass taking up the inner regions.

These density profiles, combined with the composition information we
have for these collision products, hint at what the subsequent
evolution of these stars will be. There are some interesting
trends to notice. First, the central density of the collision product
depends on the central density of the impactor, in that the more
evolved impactors produce more centrally concentrated collision
products. This is in agreement with trends suggested by earlier
simulations of collision products \citep{LRS96}. More
evolved stars have higher central densities and therefore lower
central specific entropies.  Gentle stellar collisions can be modeled
by entropy-sorting of the fluid from both stars, and so the cores of
evolved stars will end up at the centre of any collision product. The
more evolved a star is, the more of it will have low entropy and
will move to the core of the daughter star.

Another interesting trend is that the total mass of the collision
product is lower as the impactor gets more evolved. The PMS+GB
collision has a total mass of 1.38 \msun, while the PMS+PMS collision
product has a mass of 1.55 \msun. The larger mass loss with more
evolved (and hence more compact) impactors is related to the amount of
shock-heating in the collision. The PMS+WD collision product actually
loses a smaller percentage of its total mass than the PMS+GB collision
since the white dwarf as we model it does not have an atmosphere to be
shock heated during a collision. Therefore, all the mass lost during
the collision must come only from the pre-main sequence star.

A goal of this paper is to predict the evolutionary state of the
collision products after the collision, and these density profiles
will help us do that. First, the PMS+PMS collision results in a very
extended, fluffy object with no composition gradient. This object will
return to the pre-main sequence phase on a thermal timescale, and will
simply be a more massive PMS star. It will be indistinguishable from
other stars of the same mass, except that it will be (slightly)
delayed in its evolution. However, the difference in age between a 1.5
\msun~ collision product and a normal 1.5 \msun~ star in the same
cluster will be at most a few 10's of Myr, and probably not observable
except under exceptional circumstances.

The two PMS+MS collision products have denser cores than the PMS+PMS
collision product, and a steeper density gradient. These stars will
relax to become main sequence stars. As we can see from figure
\ref{composition}, some hydrogen-rich material has been mixed into the
core of the PMS+TAMS collision product, but not very much. Therefore,
these collision products can be well-modeled as `rejuvinated' stars
-- their evolution will be the same as that of a normal star of their
mass, but the starting point of these collision products is determined
by the helium content of their cores. If the main sequence impactor
was near the end of its main sequence life, then the collision product
will be near the end of its life.

Both the PMS+WD and PMS+GB collision products have `point masses' at
their core. These point masses, which are very dense, and probably
degenerate, come from either the initial giant or the entire white
dwarf. Therefore, the collision product will look very much like a
giant star -- a dense core surrounded by a hydrogen-rich
envelope. Again, the evolutionary starting point of the collision
product will be determined by the impactor. The core mass of the new
giant will be given by the core mass of the initial giant, or by the
mass of the white dwarf. In some cases, the evolution will be a little
odd -- for example, the core mass of our PMS+WD collision product is
0.6 \msun, much larger than the canonical 0.45 \msun~ required for
helium core flash in low mass stars. That collision product will
probably evolve rapidly to the horizontal branch stage, resulting in
an unusually massive HB star. Collisions between PMS and giants,
however, will result in massive giants with reasonably small core
masses. They will be slightly bluer than normal giant branch stars in
the cluster, and will also produce unusually bright and red horizontal
branch stars when they reach that stage. These could be one
explanation for the supra-horizontal branch stars seen in some
globular clusters \citep{B94,FPRD99}, for example (although we do not
expect that this is the dominant mechanism as the relevant timescales
are quite short).

For a given impactor, the amount of mass lost by the system increases
as the impact parameter decreases.  The more grazing the collision,
the less likely particles are to be given a radial velocity sufficient
to free them from the stars gravitational bound, and thus the escaper
mass is less. However, the mass lost is almost independent of velocity at
infinity. The velocity of the encounter seems to have little effect on
the mass lost during the collision.  The mass lost only increases
slightly for higher velocity collisions.  All collisions, even the
most gentle, result in impact speeds which are larger than the sound
speed in the outer layers of the PMS star.  All the gas that could
easily be shock-heated, has been heated, and has been lost.  As
expected, the total angular momentum of the system is larger for
larger impact parameters, since the angular momentum comes only from
the initial orbit.

We expect that these results are basically independent of the mass of
the pre-main sequence star. The structure of these low mass stars is
quite similar. Since they are chemically homogeneous and fully
convective, pre-main sequence stars have almost constant entropy
throughout their interior. Since they are larger and less dense, they
have more entropy than a main sequence star of the same mass. Gentle
stellar collisions result in a 'sorting by entropy' of the two parent
stars \citep{LWRSW02}. Our result that the pre-main sequence star forms
an envelope around any more evolved star during a collision is in
agreement with this general picture, and should be independent of the
masses of the pre-main sequence stars involved.

It seems clear from our simulations that dynamical models of clusters
do not need to include detailed (hydrodynamic) models of collision
products that involve pre-main sequence stars. A simple prescription
based on the structure of the impactor and assuming a mass loss of a
few percent of the total mass should suffice for most dynamical
simulations at the moment. Since pre-main sequence stars are
chemically homogeneous, they are even simpler to model than main
sequence stars. Current dynamical simulations of clusters that include
the effects of stellar collisions \citep[e.g.][]{HTAP01,PMHM01} use a
formulaic prescription for the rejuvination of main sequence stars
when they collide with each other; such a prescription could be easily
modified to include the effects of pre-main sequence stars. Efforts to
include full stellar evolution codes into stellar dynamics codes are
in their infancy, but are being spearheaded by the MODEST
collaboration \citep{MOD1}.

\section{Summary}

We have performed simulations of collisions between pre-main sequence
stars and a variety of other kinds of stars. We investigated the
effects of impact parameter and velocity of encounter, with the goal
of determining the structure and likely evolution of the collision
products. We find that pre-main sequence stars are not very disrupted
by collisions, and tend to wrap themselves around the outside of their
impactors. As the impactor becomes more evolved (and hence more
dense), the pre-main sequence star material gains more energy in the
collisions and is less likely to remain bound to the system. The mass
lost by the collision increases as the impact parameter decreases, and
is largely independent of the velocity of the encounter in the range
we studied ($v_{\rm inf} = 10 - 100$ km $s^{-1}$).

The subsequent evolution of the collision products was speculated
based on their density and composition profiles immediately after the
collision. PMS+PMS collision products should become higher mass
pre-main sequence stars. PMS+MS collisions should produce
higher mass main sequence stars with their starting point on the
evolutionary track determined by the helium content of the main
sequence impactor's core. Collisions between pre-main sequence stars
and stars with helium cores or white dwarfs should result in giant
branch stars, with their initial evolutionary point determined by the
core mass of the impactor.

\acknowledgements DL is supported in part by an NSERC Undergraduate 
Summer Research Award. AS is supported by NSERC. The simulations
reported in this paper were performed at the SHARCNet facilities at
McMaster University.

\end{document}